\documentclass[final,5p,times,twocolumn]{elsarticle}

\usepackage{amssymb}
\usepackage{amsthm}
\usepackage{bm}

\journal{Physica E}

\begin{document}

\begin{frontmatter}



\title{Atomic scale 0-$\pi$ transition in a high-$T_c$ superconductor / ferromagnetic-insulator / high-$T_c$ superconductor Josephson junction}


\author[label1,label2]{Shiro Kawabata}
\author[label3]{Yukio Tanaka}
\author[label4]{Yasuhiro Asano}
\address[label1]{Nanosytem Research Institute (NRI), National Institute of Advanced Industrial Science and Technology (AIST), Tsukuba, Ibaraki, 305-8568, Japan}
\address[label2]{CREST, Japan Science and Technology Corporation (JST), Kawaguchi, Saitama, 332-0012, Japan}
\address[label3]{Department of Applied Physics, Nagoya University, Nagoya, 464-8603, Japan}
\address[label4]{Department of Applied Physics and Center for Topological Science and Technology, Hokkaido University, Sapporo, 060-8628, Japan}

\begin{abstract}
We study the Josephson transport in a high-$T_c$ superconductor/ferromagnetic-insulator(FI)/high-$T_c$ superconductor numerically.
We found the formation of a $\pi$-junction in such systems.
More remarkably  the ground state of such junction alternates between 0- and $\pi$-states when thickness of FI
is increasing by a single atomic layer.
 We propose an experimental setup for observing the atomic-scale 0-$\pi$ transition.
Such FI-based $\pi$-junctions can be used to implement highly-coherent quantum bits.
\end{abstract}


\end{frontmatter}


\section{Introduction}
\label{}

There is an increasing interest in the novel properties of interfaces and junctions of ferromagnetic materials and superconductor~\cite{rf:Golubov,rf:Buzdin1}.
One of the most interesting effects is the formation of a  Josephson $\pi$-junction in superconductor/ferromagnetic-metal/superconductor (S/FM/S) heterostructures~\cite{rf:Buzdin2}.
In the ground-state phase difference between two coupled superconductors is $\pi$ instead of 0 as in the ordinary 0-junctions.
In terms of the Josephson relationship, $I_J= I_C \sin \phi$, where $\phi$ is the phase difference between the two superconductor layers, a transition from the 0 to $\pi$ states
implies a change in sign of $I_C$ from positive to negative.

As for the application of the $\pi$ junction, a quiet qubit consisting of a superconducting loop with a S/FM/S $\pi$-junction has been proposed~\cite{rf:Ioffe,rf:Blatter}.
In this qubit, a quantum two-level system  is spontaneously generated and therefore it is expected to be robust to the decoherence by the fluctuation of the external magnetic field.
From the viewpoint of the quantum dissipation, however,  S/FM/S junctions are identical with S/N/S junctions (N is a normal nonmagnetic metal).
Thus a gapless quasiparticle excitation in the FM layer is inevitable and  gives a strong dissipative or decoherence effect~\cite{rf:Schon,rf:Kato}.
Therefore the realization of the $\pi$-junction without a metallic interlayer is highly desired for qubit applications~\cite{rf:Kawabata1,rf:Kawabata2,rf:Kawabata3,rf:Kawabata4}.

Recently we have theoretically predicted that the $\pi$ junction can be formed in low-$T_c$ (LTSC) superconductor /  La${}_2$BaCuO${}_5$(LBCO) / LTSC junctions~\cite{rf:Kawabata5,rf:Kawabata6,rf:Kawabata7,rf:Kawabata8}.
Here LBCO is a representative material of $ferromagnetic$ $insulators$ (FIs)~\cite{rf:Mizuno}.
More remarkably the ground state of such junctions alternates between 0- and $\pi$-states when thickness of FI
is increasing by a single atomic layer.

However, in order to observe  the atomic scale  0-$\pi$ transition, we have to fabricate the junction with completely flat interface between FI and superconductors.
Therefore, from the perspectives of the FI/superconductor  interface matching,  the usage of high-$T_c$ cuprate superconductors (HTSC), e.g., YBa${}_2$Cu${}_3$O${}_{7-\delta}$ and La${}_{2-x}$Sr${}_x$CuO${}_4$ (LSCO) is desirable, because recent development of the pulsed laser deposition technique enable us to fabricate layer-by-layer epitaxial-growth of such oxide materials~\cite{rf:Mercey,rf:Prellier}.
Thus we can expect the experimental observation of the 0-$\pi$ transition by increasing the layer number of LBCO.
In this paper, we investigate the Josephson effect for  a HTSC/FI/HTSC junction theoretically and show that the atomic scale 0-$\pi$ transition can be realized in such realistic oxide-based  junctions.
We also propose an experimental setup for detecting the atomic scale 0-$\pi$ transition.

\section{Model}
\label{}

Let us consider a three-dimensional tight-binding square-lattice of a HTSC/FI/HTSC junction with $L_x$ and $L_y$ being the numbers of the lattice sites in the $x$ and $y$ directions  as shown in Fig.~1(a).
The vector 
\begin{eqnarray}
{\bf r}=x {\bf x}+y{\bf y} +z{\bf z}
\end{eqnarray}
 points to a lattice site, where ${\bf x}$ and ${\bf y}$ are unit vectors in the $x$ and $y$ directions in the HTSC plane,
respectively.
The lattice constant is set to be unity.
In the $x$ and $y$ directions, we apply the hard wall boundary condition for the number of lattice sites being $L_x=L_y \equiv M$.
Electronic states in a $d$-wave HTSC are described by the mean-field BCS Hamiltonian, 
\begin{eqnarray}
{\cal H}_\mathrm{HTSC}
=
-t
\sum_{{\bf r},{\bf r}^{\prime }, \sigma} 
c_{{\bf r} \sigma }^\dagger
c_{{\bf r}^\prime \sigma }
+
(2 t -\mu_s) 
\sum_{{\bf r} , \sigma} 
c_{{\bf r} \sigma }^\dagger
c_{{\bf r} \sigma }
\nonumber\\
+
\frac{1}{2} 
\sum_{\bf r}
\left[
 \Delta 
   c_{{\bf r}+x, \uparrow }^\dagger
   c_{{\bf r} \downarrow }^\dagger
+ 
\Delta 
   c_{{\bf r}-x, \uparrow }^\dagger
   c_{{\bf r} \downarrow }^\dagger
+
 \Delta^* 
  c_{{\bf r} \downarrow }
     c_{{\bf r}+x, \uparrow }
 +
 \Delta^* 
  c_{{\bf r} \downarrow }
     c_{{\bf r}-x, \uparrow }
     \right.
     \nonumber\\
-
     \left.
 \Delta 
   c_{{\bf r}+y, \uparrow }^\dagger
   c_{{\bf r} \downarrow }^\dagger
- 
\Delta 
   c_{{\bf r}-y, \uparrow }^\dagger
   c_{{\bf r} \downarrow }^\dagger
-
 \Delta^* 
  c_{{\bf r} \downarrow }
     c_{{\bf r}+y, \uparrow }
-
 \Delta^* 
  c_{{\bf r} \downarrow }
     c_{{\bf r}-y, \uparrow }
\right]
.
\end{eqnarray}
Here
  $
  c_{{\bf r} ,\sigma }^{\dagger }$ ($c_{{\bf r},\sigma }^{{}}
$)
 is the creation
(annihilation) operator of an electron at ${\bf r}$ with spin
 $\sigma
=$ ( $\uparrow $ or $\downarrow $ ) and $\mu_s$ is the chemical potential.
The hopping integral $t$ is considered among nearest neighbor sites and $\Delta$ is the amplitude of $d$-wave pair potential.

\begin{figure}[t]
\begin{center}
\includegraphics[width=8.6cm]{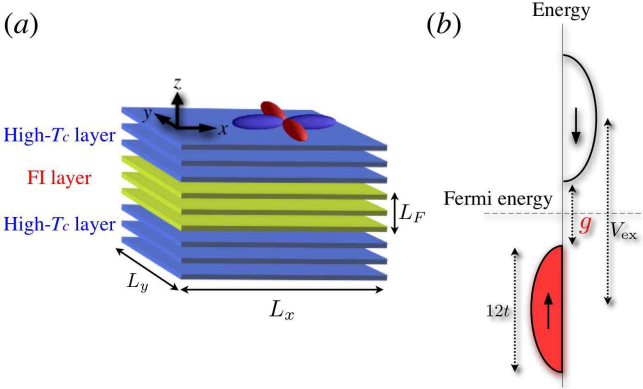}
\end{center}
\caption{(Color online) (a) A schematic figure of a hight-$T_c$ superconductor Josephson junction through the ferromagnetic-insulators on the
tight-binding lattice. The density of states for each spin direction for the  ferromagnetic-insulator, e.g., LBCO.
 }
\label{fig1}
\end{figure}

The typical DOS of FI for each spin direction is shown schematically in Fig. 1(b).
LBCO is the one of the representative of FI.
In 1990, Mizuno $et$ $al$, found that LBCO undergoes a ferromagnetic transition at 5.2 K~\cite{rf:Mizuno}.
The exchange splitting $V_\mathrm{ex}$ is estimated  to be 0.34 eV by a first-principle band calculation using the spin-polarized local density approximation~\cite{rf:LBCO}.
Since the exchange splitting is large and the bands are originally half-filled, the system becomes FI. 
The Hamiltonian of a FI layer can be described by a single-band tight-binding model~\cite{rf:Kawabata6} as 
\begin{eqnarray}
{\cal H}_\mathrm{FI}
& =& -t \sum_{{\bf r},{\bf r}^{\prime },\sigma} 
c_{{\bf r},\sigma}^\dagger 
c_{{\bf r}',\sigma}
-\sum_{{\bf r}} ( 4 t -\mu)  
c_{{\bf r},\uparrow}^\dagger 
c_{{\bf r},\uparrow}
\nonumber\\
&+&
 \sum_{{\bf r}} 
( 4 t -\mu + V_\mathrm{ex}) 
 c_{{\bf r},\downarrow}^\dagger 
 c_{{\bf r},\downarrow}
,
\end{eqnarray}
where $V_\mathrm{ex}=12 t+ g$ ($g$ is the gap between up  and down spin band) is the exchange splitting and $\mu$ is the chemical potential [see Fig. 1(b)].
If $V_\mathrm{ex} >  12 t$, this Hamiltonian describes FI as Fig. 1(b).

The Hamiltonian is diagonalized by the Bogoliubov transformation.
The Andreev bound state consists of subgap states whose
wave functions decay far from the junction interface.
In what follows, we focus on the subspace for
spin-$\uparrow$ electron and spin-$\downarrow$ hole.
In superconductors, the wave function of a bound state is given by
\begin{eqnarray}
\Psi_{l,m}^1(\boldsymbol{r})=&\Phi_1 \left[
 \left(\begin{array}{c} u \\ v \end{array} \right)
 A e^{-ikz} +
 \left(\begin{array}{c} v \\ u \end{array} \right)B e^{ik z}\right] \chi_l(x) \chi_m(y) ,
 \\
\Psi_{l,m}^2 (\boldsymbol{r})=&\Phi_2 \left[
 \left(\begin{array}{c} u \\ v \end{array} \right)
 C e^{ikz} +
 \left(\begin{array}{c} v \\ u \end{array} \right)D e^{-ikz}\right] \chi_l(x) \chi_m(y).
 \end{eqnarray}
Here $\nu=1$ ($2$) indicates an upper (lower) superconductor, $A, B, C$ and $D$
are amplitudes of the wave function for an outgoing quasiparticle, 
$\phi_\nu$ is the phase of a superconductor,
\begin{eqnarray}
\Phi_\nu&=&\mathrm{diag} \left(  e^{i\phi_\nu/2} , e^{-i\phi_\nu/2} \right)
\\
u&=&\sqrt{\frac{1}{2} \left(  1+ \frac{\Omega_{lm}}{E} \right)}
\\
v&=&\sqrt{\frac{1}{2} \left(  1- \frac{\Omega_{lm}}{E} \right)}
,
\end{eqnarray}
with $\Omega_{lm}= \sqrt{E^2 - \Delta_{lm}^2}$ and  $\Delta_{lm} 
=
\Delta \left( \cos q_l -  \cos q_m \right)  
$,
where $q_l= \pi l /(M+1)$ and  $q_m= \pi m /(M+1)$.
The wave function in the $x$ and $y$ directions is given by 
\begin{eqnarray}
\chi_l(x) &=& \sqrt{ \frac{1}{M +1}}   \sin \left( \frac{\pi}{M +1} l \right)
\\
\chi_m(y) &=& \sqrt{ \frac{1}{M +1}}   \sin \left( \frac{\pi}{M +1} m \right)
,
\end{eqnarray}
where $l$ and $m$ indicate a transport channel.
The energy $E$ is measured from the Fermi energy
and 
\begin{eqnarray}
k= \cos^{-1} 
\left(
4-\frac{\mu_s}{2t}  - \cos q_l -\cos q_m -i \frac{\sqrt{\Delta_{lm}^2 -E^2} }{ 2 t} 
\right)
\end{eqnarray}
is the complex wave number.
In a FI, the wave function is given by
\begin{eqnarray}
\Psi_\mathrm{FI}(\boldsymbol{r})&=&
\left[ \left(\begin{array}{c} f_1 e^{-iq_e z} \\ g_1 e^{-iq_h z}\end{array} \right)
+\left(\begin{array}{c} f_2 e^{iq_e z} \\ g_2 e^{iq_h z}\end{array} \right)\right] \chi_l(x) \chi_m(y)
,
\end{eqnarray}
with
\begin{eqnarray}
 q_e&=&\pi + i\beta_\uparrow, \label{wn1}\\
   q_h&=&i\beta_\downarrow,\label{wn2}
\end{eqnarray}
where 
\begin{eqnarray}
 \cosh\beta_\uparrow
 &=&
 1+ \frac{E}{2 t} + \frac{g}{4t} + \cos q_l + \cos q_m - 2 \cos \left( \frac{\pi M }{M+1} \right)
 \nonumber\\
 \\
 \cosh \beta_\downarrow
  &=&
 1+ \frac{E}{2 t} + \frac{g}{4t} - \cos q_l - \cos q_m - 2 \cos \left( \frac{\pi  }{M+1} \right)
 \nonumber\\
  \end{eqnarray}
and
$f_1, f_2, g_1$ and $g_2$ are amplitudes of wave function in a FI.
The Andreev levels $\varepsilon_{n,l,m} (\phi=\phi_L-\phi_R)$ [$n=1, \cdots, 4$]
can be calculated from boundary conditions
\begin{eqnarray}
\Psi_1(x,y,\lambda)&=&\Psi_\mathrm{FI}(x,y,\lambda)
\\
\Psi_2(x,y,L_F+\lambda)&=&\Psi_\mathrm{FI}(x,y,L_F+\lambda)
  \end{eqnarray}
 for $\lambda =0$ and 1.
The Josephson current is related to $\varepsilon_{n,l}$ via
\begin{eqnarray}
 I_J (\phi) =\frac{2e }{ \hbar} \sum_{n,l,m}  
 \frac{\partial \varepsilon_{n,l,m} (\phi)}
{ \partial \phi }
 f  \left[ \varepsilon_{n,l,m} (\phi) \right]
 ,
  \end{eqnarray}
where $ f\left( \varepsilon  \right)$ is the Fermi-Dirac distribution function.
In the case of a high barrier limit ($g \gg t$) which is appropriate for LBCO, the Josephson current phase relation is given by $I_J(\phi) = I_C \sin \phi$.
Thus we define the Josephson critical current $I_C$ as  $I_C = I_J(\pi/2)$.

\section{Numerical results}
\label{}

In this section, in order to show the possibility of $\pi$-coupling in such realistic HTSC junctions, we numerically calculate the Josephson critical current $I_C$ [Fig. 2].
 The tight binding parameters $t$ and $g$  have been determined  by fitting to the first-principle band structure calculations~\cite{rf:LBCO} as $g/t=20$.
Figure 2 shows the FI thickness $L_F$ dependence of $I_C$ at $T=0.01 T_c$  ($T_c$ is the superconducting transition temperature) for a LSCO/LBCO/LSCO junction with $V_\mathrm{ex}/t=32$, $\Delta_d/t=0.6$, and $M=L_x=L_y=100$.
As expected, the atomic scale  0-$\pi$ transitions can be realized in such oxide-based junctions.
The physical origin of this transition can be explained by the thickness-dependent phase shifts between the wave numbers of electrons and holes in FIs as in the LTSC junctions~\cite{rf:Kawabata6}.

It is important to note  that in the case of stack HTSC Josephson junctions~\cite{rf:Kleiner,rf:Yurgens}, no zero-energy Andreev bound-states~\cite{rf:Kashiwaya} which give a strong Ohmic dissipation~\cite{rf:KawabataMQT1,rf:KawabataMQT2,rf:KawabataMQT3} are formed.
Moreover, the harmful influence of nodal-quasiparticles due to the $d$-wave order-parameter symmetry on the macroscopic quantum dynamics in such  junctions is found to be weak both theoretically~\cite{rf:Fominov,rf:Amin,rf:KawabataMQT4,rf:Umeki,rf:KawabataMQT5} and experimentally~\cite{rf:InomataMQT,rf:Jin,rf:Matsumoto,rf:KashiwayaMQT}.
Therefore HTSC/LBCO/HTSC $\pi$-junctions would be a promising candidate for quiet qubits.

\begin{figure}[b]
\begin{center}
\includegraphics[width=8.8cm]{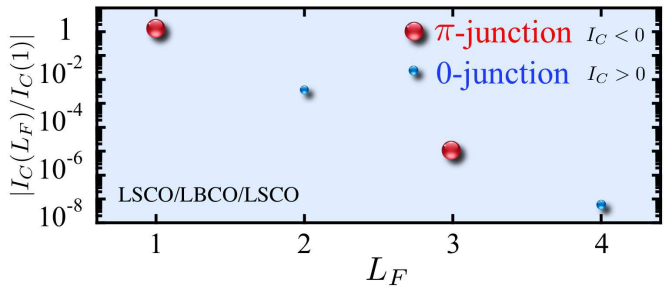}
\end{center}
\caption{(Color online) The Josephson critical current $I_C$ as a function of the FI thickness $L_F$ at $T=0.01 T_c$ for a $c$-axis stack LSCO/LBCO/LSCO junction with $V_\mathrm{ex}/t=28$, $\Delta_d/t=0.6$, and $M=L_x=L_y=100$.
The large red (small blue) circles indicate the $\pi$(0)-junction.}
\label{fig5}
\end{figure}
%
%
%
%
%
%
%
%

\section{Experimental setup for observing the $\pi$ junction behavior}
\label{}

We would like to show an experimental setup for observing the $\pi$ junction and the atomic scale  0-$\pi$ transition.
The formation of the $\pi$-junction can be experimentally detected by using a HTSC ring [see Fig. 3].
The phase quantization condition for the HTSC ring is given by 
 \begin{eqnarray}
 2 \pi \frac {\Phi -\Phi_\mathrm{ext} }{ \Phi_0} + \phi_1 + \phi_2 = 2 \pi n
,
  \end{eqnarray}
 where
$\phi_1$ and $\phi_2$ are  the phase difference across the junctions 1 and 2, $\Phi$ is the magnetic flux penetrating through the ring, $\Phi_0$ is the flux quantum, and $n$ is an integer.
The current passed through the ring divides between the junctions 1 and 2, $i.e.,$ 
 \begin{eqnarray}
  I=I_{C1} \sin \phi_1 +I_{C2} \sin \phi_2
.
  \end{eqnarray}
Applied external magnetic flux $\Phi_\mathrm{ext}$ depletes phases $ \phi_1$ and $ \phi_2$ causing interference between currents through the junctions 1 and 2.
For a symmetric ring with $I_{C1} \approx I_{C2} = I_C$ and negligible geometric inductance ($L =0$), the total critical current as a function of $\Phi_\mathrm{ext}$  is given by 
 \begin{eqnarray}
 I_C^{00} =I_C^{\pi \pi}=2 I_C \left|  \cos \left( \pi \frac{ \Phi_\mathrm{ext} }{ \Phi_\mathrm{0} } \right) \right| 
,
  \end{eqnarray}
for the case that $L_F$ of the both junctions are same.
If $L_F$ of the junction 1(2) is even and $L_F$ of the junction 2(1) is odd, we get
 \begin{eqnarray}
I_C^{0 \pi} =I_C^{\pi 0}=2 I_C \left|  \sin \left( \pi \frac{ \Phi_\mathrm{ext} }{ \Phi_\mathrm{0} } \right) \right| 
.
  \end{eqnarray}
Therefore the critical current of a 0-$\pi$ (0-0) ring has a minimum (maximum) in zero applied magnetic field~\cite{rf:Sigrist}.
Experimentally, the half-periodic shifts in the interference patterns of the HTSC ring can be used as a strong evidence of the $\pi$-junction and also the atomic scale 0-$\pi$ transition.
Such a half flux quantum shifts have been observed in a $s$-wave ring made with an LTSC/FM/LTSC~\cite{rf:Guichard} and a LTSC/quantum dot/LTSC junction~\cite{rf:Dam}.

\begin{figure}[t]
\begin{center}
\includegraphics[width=8.0cm]{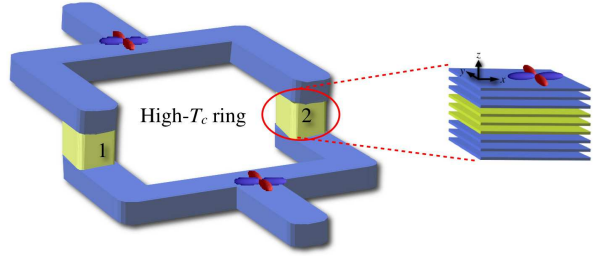}
\end{center}
\caption{(Color online) Schematic picture of the high-$T_c$ ring which can be used in experimental observations of  the $\pi$-junction and the atomic scale 0-$\pi$ transition.}
\label{fig3}
\end{figure}
%
%
%
%
%

\section{Summary}
\label{}

To summarize, we have studied the Josephson effect in HTSC/FI/HTSC junctions by use of the three-dimensional tight-binding model.
We found that the $\pi$-junction and the atomic-scale 0-$\pi$ transition can be realized in realistic junctions.
Such FI based $\pi$-junctions can be used as an  element in the architecture of  {\it ideal quiet qubits} which possess both the quietness and the weak quasiparticle-dissipation nature.
Therefore, ultimately, we could realize a FI-based highly-coherent quantum computer.

\section*{Acknowledgement}
\label{}
We  would like to thank J. Arts, A. Brinkman, M. Fogelstr\"om, A. A. Golubov, S. Kashiwaya,  P. J. Kelly, T. L\"ofwander, F. Nori, and M. Weides for useful discussions.
This work was  supported by CREST-JST, and a Grant-in-Aid for Scientific Research from the Ministry of Education, Science, Sports and Culture of Japan (Grant No. 22710096).





\bibliographystyle{elsarticle-num}
\bibliography{<your-bib-database>}







\end{document}